\documentclass[useAMS,usenatbib,usegraphicx]{mn2e}
\def\beq{\begin{equation}}
\def\eeq{\end{equation}}
\def\beqa#1{\begin{eqnarray}\label{#1}}
\def\eeqa{\end{eqnarray}}

\title{ The First Stars in The Universe}

\author[S. Naoz, S. Noter and R. Barkana]{S. Naoz, S. Noter and R. Barkana$^{1}$
\thanks{E-mail: smadar@wise.tau.ac.il (SN); shaynote@post.tau.ac.il (SN); 
barkana@wise.tau.ac.il (RB)}\\ $^{1}$School of Physics and Astronomy,
The Raymond and Beverly Sackler Faculty of Exact Sciences,\\ Tel Aviv
University, Tel Aviv 69978, ISRAEL}

\begin{document}

\pagerange{\pageref{firstpage}--\pageref{lastpage}} \pubyear{2006}

\maketitle

\label{firstpage}

\begin{abstract}
Large telescopes have allowed astronomers to observe galaxies that
formed as early as 850 million years after the Big Bang. We predict
when the first star that astronomers can observe (i.e., in our past
light cone) formed in the universe, accounting for the first time for
the size of the universe and for three essential ingredients: the
light travel time from distant galaxies, Poisson and density
fluctuations on all scales, and the effect of very early cosmic
history on galaxy formation. We find that the first observable star is
most likely to have formed 30 million years after the Big Bang (at
redshift 65). Also, the first galaxy as massive as our own Milky Way
likely formed when the universe was only 400 Myr old (at redshift
11). We also show that significant modifications are required in
current methods of numerically simulating the formation of galaxies at
redshift 20 and above.
\end{abstract}

\begin{keywords}
galaxies:high-redshift -- cosmology:theory -- galaxies:formation
\end{keywords}

\section{Introduction}

The formation of the first stars, in a universe that had not yet
suffered chemical and magnetic stellar feedback, is a well-defined
problem for theorists \citep{review}. The advent of large telescopes
has accelerated the observational quest for the first objects as
astronomers have detected the light emitted by increasingly distant
galaxies \citep{fan03}. A theoretical understanding of the first stars
is of great importance as observations approach the pristine regime of
the early universe.

Stars are understood to form at high redshift out of gas that cooled
and subsequently condensed to high densities in the cores of dark
matter halos. Since metals are absent in the pre-stellar universe, the
earliest available coolant is molecular hydrogen ($H_2$), and thus the
minimum halo mass that can form a star is set by requiring the
infalling gas to reach a temperature $\ga 1000$ K required for
exciting the rotational and vibrational states of $H_2$
\citep{th2}. This has been confirmed with high-resolution numerical
simulations containing gravity, hydrodynamics, and chemical processes
in the primordial gas \citep{abel,bromm}. These simulations showed
that the first star formed within a halo containing $\sim 10^5
M_{\odot}$ in total mass; however, as we show, they could not estimate
{\it when}\/ the first star formed.

\section{Limitations of simulations}

In the real universe, the cosmological redshift $z$ of each source
implies a corresponding cosmic age at which the source is seen. Thus,
the observable universe can be divided into thin spherical shells
around us, each containing a portion of the universe that is observed
at a given redshift $z$ and is seen at the corresponding age. Galaxy
formation in different shells is correlated due to density
fluctuations on various scales. A simulation that wishes to determine
when various objects are observed to form must encompass a sufficient
portion of our past light cone.

Even with adaptive grid codes, full simulations are limited by current
technology to tiny regions and only form a first star below redshift
20 \citep{abel} or 33 for the most recent simulations \citep{ON}. Even
when great sacrifices are made in the realism, so that hydrodynamics
and chemical evolution are neglected and the collapse of a star is
only partially resolved, simulations still cannot capture the proper
cosmological context. A direct simulation of the universe out to the
spherical shell at redshift 70 would require a simulated box of length
$25,000$~Mpc on a side [measured in comoving distance, which equals
physical distance times a factor of $(1+z)$]. The need to resolve each
$10^5 M_{\odot}$ halo into $500$ particles (required to determine the
halo's evolution even crudely \citep{Springel03}) determines the mass
of each simulated particle, while the mean cosmic density of $4\times
10^{10} M_\odot/$Mpc$^3$ yields the total mass of the simulated box,
and implies a required total of $3\times10^{21}$ particles. The
maximum number achieved thus far is only $\sim 10^{10}$
\citep{Springel}. Early times have been probed by resimulating regions
from a large initial simulation \citep{reed}, forming star-forming
halos at $z=47$ (the earliest to date), but with cosmological
parameters for which the first galaxy actually formed at $z \sim 82$
according to our model below. Thus, the actual starting redshift of
the era of galaxy formation can currently only be predicted
analytically. For similar parameters, \citet{jordi} made a rough
analytical estimate (using the
\citet{press} halo mass function) and found that the first star formed 
at $z\sim 48$. A rough estimate by \citet{wa} found $z\sim 71$ using
the improved \citet{Sheth99} mass function (see also \S~3).

Simulations of the first stars also face a difficulty in the way their
initial conditions are determined. Recent observations of the cosmic
microwave background (CMB) \citep{spergel06} have confirmed the notion
that the present large-scale structure in the universe as well as the
galaxies within it originated from small-amplitude density
fluctuations at early cosmic times. Simulations thus assume Gaussian
random initial fluctuations as might be generated by a period of
cosmic inflation in the early universe. The evolution of these
fluctuations can be calculated exactly as long as they are small, with
the linearized Einstein-Boltzmann equations. The need to begin when
fluctuations are still linear forces numerical simulations of the
first star-forming halos to start at a high redshift, the highest
to-date being 600 \citep{reed}. Current simulations neglect the
contribution of the radiation to the expansion of the universe and
assume the Newtonian limit of general relativity which is valid in
simulation boxes that are small with respect to the horizon. However,
for a halo that forms around redshift 65, we find below that the
corresponding region already must have had a moderately non-linear
overdensity of $\delta \sim 20\%$ at $z=600$ ($\delta$ is the density
perturbation divided by the cosmic mean density). Indeed, we find that
for initial conditions that ensure $1\%$ accuracy in $\delta$,
simulations must start deep in the radiation dominated era, at $z >
10^6$. Furthermore, we show that a small error in $\delta$ leads to a
much larger error in the abundance of halos at each redshift. The
non-linearity of the initial conditions represents a
previously-unrecognized fundamental limitation of current methods of
simulating the formation of the first star-forming halos.

\section{Spherical collapse}

While numerical simulations cannot determine the formation time of the
first observable star, galaxy and halo formation can also be
understood and described with a standard analytical model that has
been quantitatively checked with simulations where the latter are
reliable. The standard analytical model for the abundance of halos
\citep{press,bond} considers the linear density fluctuations at some
early, initial time, and attempts to predict the number of halos that
will form at some later time corresponding to a redshift $z$. First,
the fluctuations are evolved to the redshift $z$ using the equations
of linear fluctuations, i.e., they are artificially linearly
extrapolated even if they may become non-linear in some regions. The
average density is then computed in spheres of various sizes. Whenever
the overdensity in a sphere containing mass $M$ rises above a critical
threshold $\delta_c(M,z)$, the corresponding region is assumed to have
collapsed by redshift $z$, forming a halo of at least mass $M$. Thus,
the model separates the linear growth of fluctuations, which we
calculate with standard codes \citep{cmbfast,NB,naoshi}, from the
non-linear collapse of objects, which determines the effective linear
threshold $\delta_c(M,z)$.

Within this model, the non-linear collapse of a halo is analyzed by
considering a uniform, spherically symmetric fluctuation whose
collapse can be calculated by numerically solving ordinary
differential equations; $\delta_c(M,z)$ is then determined as the
value of the linearly-extrapolated $\delta$ of the spherical region at
the moment when the actual, non-linear $\delta$ diverges (i.e., the
region collapses to a point). The classical calculation of spherical
collapse in an Einstein de-Sitter (EdS) universe (i.e., a flat
matter-dominated universe) yields $\delta_c=1.686$, a value that is
independent of the halo mass and the collapse redshift \citep{gg}.
Given the power spectrum of the initial fluctuations, together with
their linear growth, we can determine the chance that regions
containing various initial masses $M$ will reach the threshold
$\delta_c$ at some final redshift $z$, and this forms the basis for
estimating the abundance of halos of mass $M$ that form at a specific
$z$ \citep{press,bond}. We use the recently-modified form of this
model [\citet{Sheth99}, with the updated parameters suggested by
\citet{Sheth02}], which fits halo abundances in numerical simulations
very accurately in regimes that include $M=10^{11}$--$10^{15}
M_{\odot}$ halos at $z=0$--10 \citep{Jenkins,Springel} and $M=10^6
M_{\odot}$ halos at $z=15$--30 \citep{BL04,Yoshida}. We note that
other suggested mass functions \citep[e.g.,][]{reed06} are in
agreement with the \citet{Sheth99} mass function if the updated
parameters are used. Based on the plausible physical arguments behind
this model together with its quantitative success out to redshift 30,
we extrapolate it to predict halo abundances at still-higher
redshifts.

Our results assume cosmological parameters matching the latest CMB and
weak lensing observations \citep{spergel06}. For the contributions to
the energy density, we assume ratios relative to the critical density
of $\Omega_m=0.299$, $\Omega_\Lambda=0.701$, and $\Omega_b=0.0478$,
for matter, cosmological constant, and baryons, respectively. We also
assume a Hubble constant $H_0=100 h \mbox{ km s}^{-1}\mbox{Mpc}^{-1}$
with $h=0.687$, and a primordial power-law power spectrum with
spectral index $n=0.953$ and $\sigma_8=0.826$, where $\sigma_8$ is the
root-mean-square amplitude of mass fluctuations in spheres of radius
$8\ h^{-1}$ Mpc.

Within this analytical model, we can account for early cosmic history
when predicting the abundance of high-redshift stars and galaxies. We
first calculate the correct value of $\delta_c(M,z)$ with a spherical
collapse calculation that includes the full formation history of each
halo (Figure~1), starting out with a small perturbation and following
its gravitational evolution until it collapses. We begin our
calculation at a very high redshift when $\delta \ll 1$, which implies
that the overdense region containing the mass $M$ is larger than the
horizon. In this regime we use the cosmological Friedmann equations of
general relativity, applied to the overdense region that we are
following. We begin in the radiation-dominated era, and include
throughout the contribution of the radiation to the cosmic
expansion. Once the fluctuation enters the horizon, the overdense dark
matter continues to collapse due to gravity, but the radiation
pressure suppresses the sub-horizon perturbation in the photon
density, and the coupling of the baryons to the photons (via Compton
scattering) keeps the baryon perturbation small as well. Thus, in this
regime we neglect the perturbation in the radiation and baryons and
continue to evolve the spherical collapse of the dark matter
perturbation. The formation of hydrogen at cosmic recombination ($z
\sim 1200$) decouples the cosmic gas from its mechanical drag on the
CMB, and the baryons (which constitute a significant $17\%$ of all the
matter) subsequently begin to fall into the pre-existing gravitational
potential wells of the dark matter. In this regime we evolve the
coupled collapse of the dark matter and the baryons, both evolving
under their mutual gravity but with different initial conditions. We
find that a halo destined to contain the first observable star in the
universe reaches a $\delta \sim 1\%$ at $z \sim 10^6$ and grows to
$\delta \sim 6\%$ at matter-radiation equality and $\delta \sim 13\%$
at cosmic recombination. Since the fluctuation enters the horizon very
early in the radiation-dominated era, the final value of
$\delta_c(M,z)$ for a given $M$ and $z$ is insensitive to the precise
starting redshift (as long as it is much higher than equality) or the
precise treatment of the fluctuation's horizon crossing.

\begin{figure}
\includegraphics[width=84mm]{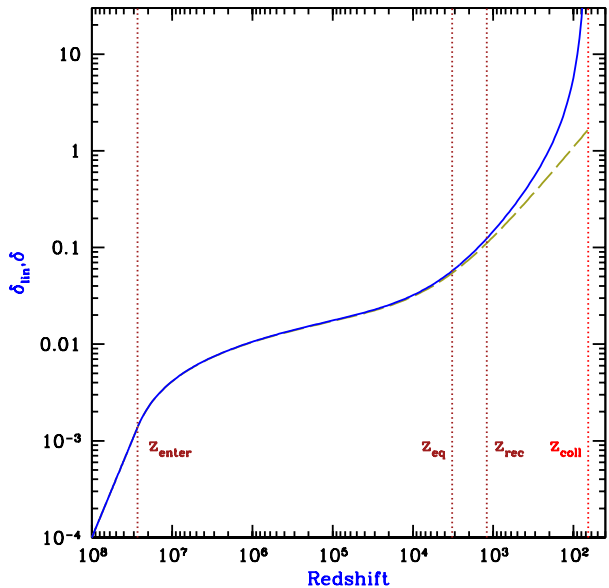}
\caption{Evolution of the fractional overdensity $\delta$ for a
spherical region containing $10^5 M_{\odot}$ that collapses at
$z=66$. We show the fully non-linear $\delta$ (solid curve) and the
linearly-extrapolated $\delta$ (dashed curve). We indicate the
redshifts of halo collapse ($z_{\rm coll}$), cosmic recombination
($z_{\rm rec}$), matter-radiation equality ($z_{\rm eq}$), and entry
into the horizon ($z_{\rm enter}$).}
%\label{Pkfig} 
\end{figure}

We find that $\delta_c$ is essentially independent of $M$, and for our
assumed parameters it is lower than the EdS value by $\sim 1\% \times
(1+z)/20$ in the range $z=9$--80. Adding the contribution of the
radiation to the expansion of the universe, as well as adding the
baryon-photon coupling to the collapse, both reduce the extrapolated
linear perturbation at collapse $\delta_c$ compared with the EdS
value. These added physical effects reduce the collapse efficiency
since only the dark matter component takes part in the collapse
throughout. The radiation is kept smooth by its own pressure and it
never participates in the collapse, while the baryons only gradually
recover from their coupling to the photons and begin to catch up to
the dark matter; during this recovery stage, as long as the baryon
perturbations remain much smaller than those of the dark matter, the
baryons essentially do not participate in the collapse. Now, any
reduction in the matter fraction that collapses (compared to $100\%$
in the usually-assumed case of pure cold dark matter) depresses the
linear evolution of the density perturbation more strongly, while the
non-linear perturbation is larger and is thus less affected by the
components that do not help in the collapse. Therefore the linear
perturbation reaches a lower value of $\delta_c$ when the non-linear
perturbation collapses. Note that while the change in $\delta_c$ may
seem small, when dealing with rare halos, even a change of a few
percent in $\delta_c$ can change the halo abundance at a given
redshift by an order of magnitude (Figure~2).  Even at redshift 20,
the $1\%$ change in $\delta_c$ compared to its previously-assumed
value changes the cumulative halo mass function $n(>M)$ by $\ga 10\%$.

\begin{figure}
\includegraphics[width=84mm]{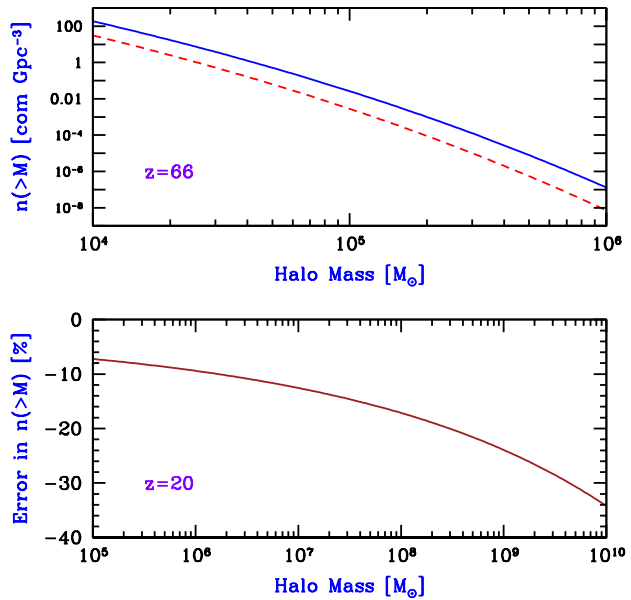}
\caption{Cumulative halo mass function $n(>M)$. At $z=66$ (top panel) 
we compare our full result for $n(>M)$ (solid curve) to that resulting
from adopting the EdS value of $\delta_c=1.686$ (dashed curve). At
$z=20$ (bottom panel), we show the relative error in $n(>M)$ that
results from using $\delta_c=1.686$ instead of our full result.}
%\label{Pkfig} 
\end{figure}

\section{Statistical considerations}

As noted earlier, the minimum halo mass that can form a star at high
redshift is determined by $H_2$ cooling \citep{th2}. Numerical
simulations \citep{abel,bromm,fuller,Yoshida,reed} imply a minimum
circular velocity $V_c \sim 4.5$ km/s, where $V_c=\sqrt{GM/R}$ in
terms of the halo virial radius $R$. This yields a corresponding
minimum mass at each redshift $z'$, $M_{\rm min}(z')$. The simulations
show that in a halo above the minimum mass, the gas cools in the dense
center and forms at least one star very quickly; this is understood
theoretically since both the cooling time and the dynamical time are a
small fraction of the cosmic age at that time.

Given the comoving number density of such star-forming halos at each
redshift, $n(M_{\rm min}(z'),z')$, the mean expected number of
star-forming halos observed at redshift $z$ or higher is \beq \langle
N \rangle (z)= \int_z^{z_{\rm max}} n(M_{\rm min}(z'),z')\, \frac{dV}
{dz'}\, dz'\ , \eeq where $dV$ is the comoving volume of the spherical
shell in the redshift range $z'$ to $z'+dz'$. Our results are
independent of the adopted maximum redshift as long as $z_{\rm
max}>80$. Although the mean expected number $\langle N \rangle (z)$ is
determined theoretically by the cosmological parameters, in practice
our observable universe presents us with a single instance and the
actual observed redshift is subject to Poisson and density
fluctuations.

The effect of Poisson fluctuations is easily calculated. The chance of
observing at least one star-forming halo above redshift $z$ with
Poisson fluctuations included is $1-\exp[-\langle N \rangle (z)]$,
which yields a spread in the redshift of the first such halo of
$\Delta z \approx 3$ at $1-\sigma$. In addition, the effect of density
fluctuations on all scales is included statistically in our analytical
model for the mean halo abundance at each redshift, but the
correlations in the halo abundance between nearby points are not
properly included. In other words, our calculation assumes that the
relevant redshift shell gives us a fair sample of the density
fluctuations on all scales. We first note that theoretically we expect
that if we included the fully correlated density field, the effect on
the predicted redshift of the first star-formation halo would be much
smaller than the spread we find due to Poisson fluctuations. Since the
universe is homogeneous on large scales, large-scale modes can shift
the halo abundance $\langle N \rangle (z)$ only by a small $\Delta z$;
indeed, only 10 Mpc scales and smaller can produce a $\Delta z > 1$
\citep{BL04}. Such small-scale modes are indeed very well sampled
within the spherical redshift shells that have a very large radius of
$\sim 12,500$ Mpc. We have also verified this with a direct
quantitative test. We have used the power spectrum to produce
instances of the full, correlated density field in three-dimensional
boxes. We have then modified the abundance $n(M_{\rm min}(z),z)$,
increasing it in high-density regions and decreasing it in low-density
regions according to a model that fits the results of numerical
simulations \citep{BL04}. In a 1 Gpc$^3$ box resolved into $256^3$
cells, we found that including a correlated density field shifts the
typical redshift of the first star-forming halo in the box by a
$\Delta z < 1$, which is already a smaller effect than the Poisson
redshift spread. The spherical shells in the real universe have a
larger volume by a factor of $\sim 2000$, so the density fluctuations
are far better sampled than in our 1 Gpc$^3$ box, and the effect of
the correlations on the redshift of the first star-forming halo in the
universe is negligible.

\section{Predictions}

The most likely redshift for the first observable star is 65.4
(Figure~3), while the median redshift at which there is a $50\%$
chance of seeing the first star is 65.8, which corresponds to a cosmic
age of $t=31$ Myr, less than a quarter of one percent of the current
age of 13.7 Gyr. The $1-\sigma$ ($68\%$) range is $z=64.7$--67.3 (or
$t=30$--32 Myr) and the $2-\sigma$ ($95\%$) range is
$z=63.9$--69.4. Note that even an error in the minimum $V_c$ as large
as a factor of 1.5 would cause only a $9\%$ change in the expected
formation redshift. Note also that in terms of the total cosmic mass
fraction contained in such halos, the first star-forming halo
corresponds to an $8.3-\sigma$ density fluctuation on the mass scale
of $10^5 M_{\odot}$. Future simulations will allow us to check the
validity of extrapolating the \citet{Sheth99} halo mass function to
such rare fluctuations.

\begin{figure}%[th]
\includegraphics[width=84mm]{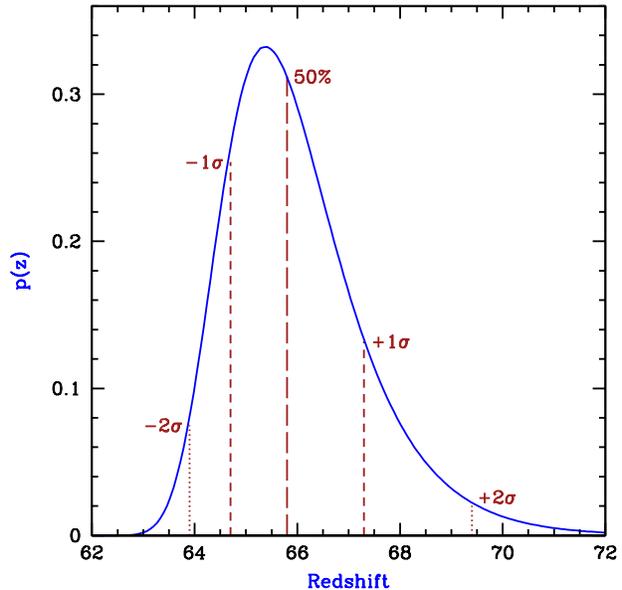}
\caption{The probability of finding the first star as a
function of redshift. Vertical lines show the median redshift as well
as the central $1-\sigma$ $(68\%)$ range and $2-\sigma$ ($95\%$)
range.}
%\label{Pkfig} 
\end{figure}

We can consider a similar question for other populations of halos
(Figure~4). For example, the radiation from the first stars is
expected to eventually dissociate all the $H_2$ in the intergalactic
medium, leading to the domination of a second generation of larger
galactic halos where the gas cools via radiative transitions in atomic
hydrogen and helium \citep{haiman}. Atomic cooling occurs in halos
with $V_c > 16.5$ km/s, in which the infalling gas is heated above
10,000 K and is ionized. We find that the first galaxy to form via
atomic cooling formed at $z=46.6^{+1.2}_{-0.9}$, which corresponds to
$t=52^{+1}_{-2}$ Myr, where we have indicated the median values and
$1-\sigma$ ranges. We also predict that the first halo as large as
that of our own Milky Way galaxy \citep{MW} formed at
$z=11.1^{+0.5}_{-0.2}$, or $t=408^{+15}_{-20}$ Myr. Also, the first
halo with the mass of the Coma galaxy cluster \citep{Coma} formed at
$z=1.24^{+0.14}_{-0.10}$, or $t=5.0\pm0.4$ Gyr.

\begin{figure}%[th]
\includegraphics[width=84mm]{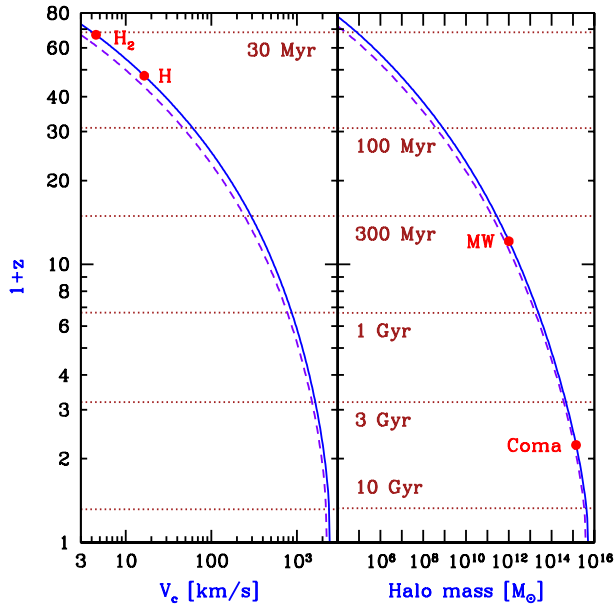}
\caption{The median redshift for the first appearance of 
various populations of halos. We consider halos above a minimum
circular velocity (left panel) or minimum mass (right panel). We
indicate in particular the first star-forming halo in which $H_2$
allows the gas to cool, the first galaxy that forms via atomic cooling
($H$), as well as the first galaxy as massive as our own Milky Way and
the first cluster as massive as Coma. The horizontal lines indicate
the elapsed time since the Big Bang. We compare the results with the
cosmological parameters from \citet{spergel06} [solid curve] to those
with the parameters from \citet{Viel} [dashed curve]; the difference
represents roughly the systematic $1\sigma$ error due to current
uncertainties in the values of the cosmological parameters.}
%\label{Pkfig} 
\end{figure}

\section{Discussion}

There remains a $\sim 10\%$ systematic uncertainty in the above $1+z$
values due to current uncertainties in the cosmological
parameters. Varying the cosmological parameters has a negligible
effect on the critical overdensity ($\delta_c$) but significantly
changes the power spectrum and also the spatial scale corresponding to
a given halo mass or circular velocity. We illustrate the
uncertainties in Figure 4 using the alternate cosmological parameters
from \citet{Viel}: $\Omega_m=0.253$, $\Omega_\Lambda=0.747$,
$\Omega_b=0.0425$, $h=0.723$, $n=0.957$ and $\sigma_8=0.785$. These
parameters also represent typical $1-\sigma$ errors, in terms of the
parameters uncertainties given by \citet{spergel06}. We obtained for
these alternate parameters that the first star formed at $z=60.0$,
which represents a $9\%$ reduction in $1+z$ compared to our standard
parameters; also the value of $1+z$ is lowered by $9\%$ for the first
atomic-cooling halo, $7\%$ for the first Milky Way-mass halo, and
$8\%$ for the first Coma-mass cluster.

On the other hand, astrophysical uncertainties about metallicity and
feedback can only affect the precise properties of a halo of a given
mass, but not whether or not it forms. Also, in a halo with a short
cooling time, processes within the halo should not induce a
significant uncertainty in the formation time of the first stars
within it, since a halo virializes at a density of $\sim 180$ times
the cosmic mean density, and so the gravitational dynamical time
within it is always smaller than the age of the universe at the time
by at least an order of magnitude. Since the baryonic component
collapses by another factor of $\sim 20$ until the collapse is halted
by angular momentum (assuming a typical halo spin parameter of $\sim
5\%$; see, e.g., \citet{mmw}), the dynamical time within a disk or a
star-forming region is likely to be shorter still by another two
orders of magnitude.

In summary, by accounting for the size of the observable universe as
well as the early cosmic history of individual halos (Figure~1), our
results extend the era of star formation much earlier than current
simulations suggest. An analytical model of galaxy formation that has
been normalized to simulation results allows us to predict the
earliest time that various halo populations can be observed
(Figure~4). Since the statistical uncertainty in this time is
dominated by Poisson fluctuations, we can analytically predict its
full probability distribution (Figure~3). Our results also show that
in order to numerically simulate galaxy formation accurately at $z \ge
20$, current simulation methods must be revised to include the effect
on a forming halo of the early cosmic history of radiation and
baryons.  Simulations that neglect this, even if they overcome the
statistical challenges by covering the volume of the entire observable
universe, will underestimate the number of halos at $z=20$ (Figure~2)
by $9\%$ for all halos above the $H_2$ cooling threshold ($6 \times
10^5 M_{\odot}$ at $z=20$), and by $15\%$ for all halos above the
atomic cooling threshold ($3 \times 10^7 M_{\odot}$ at $z=20$).

\section*{Acknowledgments}
The authors acknowledge support by Israel Science Foundation grant
629/05 and U.S. - Israel Binational Science Foundation grant 2004386.

\bsp

\label{lastpage}

\end{document}